\documentclass[letterpaper,english,reprint, aps]{revtex4-1}
\usepackage[T1]{fontenc}
\usepackage[latin9]{inputenc}
\usepackage{color}
\usepackage{amsmath}
\usepackage{amssymb}
\usepackage{graphicx}
\usepackage{setspace}

\makeatletter

\pdfpageheight\paperheight
\pdfpagewidth\paperwidth

\makeatother

\usepackage{babel}
\begin{document}

\title{The heat and work of quantum thermodynamic processes with quantum
coherence}

\author{Shan-He Su$^{1,2}$}

\author{Jin-Fu Chen$^{1,3}$}

\author{Yu-Han Ma$^{1,3}$}

\author{Jin-Can Chen$^{2}$}

\author{Chang-Pu Sun$^{1,3}$}
\email{cpsun@csrc.ac.cn}

\address{$^{1}$Beijing Computational Science Research Center, Beijing 100084,
China.~\\
$^{2}$Department of Physics, Xiamen University, Xiamen 361005, China.~\\
$^{3}$Graduate School of Chinese Academy of Engineering Physics,
Beijing 100084, China.}
\begin{abstract}
\begin{singlespace}
Energy is often partitioned into heat and work by two independent
paths corresponding to the change in the eigenenergies or the probability
distributions of a quantum system. The discrepancies of the heat and
work for various quantum thermodynamic processes have not been well
characterized in literature. Here we show how the work in quantum
machines is differentially related to isochoric, isothermal, and adiabatic
processes. We prove that the energy exchanges during the quantum isochoric
and isothermal processes are simply depending on the change in the
eigenenergies or the probability distributions. However, for a time-dependent
system in a non-adiabatic quantum evolution, the transitions between
the different quantum states representing the quantum coherence can
affect the essential thermodynamic properties, and thus the general
definitions of the heat and work should be clarified with respect
to the microscopic generic time-dependent system. By integrating the
coherence effects in the exactly-solvable dynamics of quantum-spin
precession, the internal energy is rigorously transferred as the work
in the thermodynamic adiabatic process. The present study demonstrates
that quantum adiabatic process is sufficient but not necessary for
thermodynamic adiabatic process.
\end{singlespace}
\end{abstract}
\maketitle

\section{INTRODUCTION}

Thermodynamics is reputed to have the ability to deal with the energy
transfer and work utilizing an extremely small number of variables.
Numerous quantum devices that are related to harnessing energy focus
on pioneering concepts and unexplored mechanisms. They are exemplified
by artificial photosynthesis \citep{key-1}, quantum information heat
engines \citep{key-2,key-3,key-4}, quantum thermodynamic cycles \citep{key-5,key-6},
and photovoltaic cells \citep{key-7,key-8}. The design of an improved
quantum-inspired energy converter compels us to clarify the heat exchange
and work done from microscopic mechanisms, and extend classical thermodynamic
processes to quantum-mechanical systems.

According to the first and second laws of thermodynamics, the infinitesimal
internal energy variation of a quantum system is
\begin{equation}
dU=\bar{d}Q+\bar{d}W=\sum_{n}\left(E_{n}d\rho_{nn}+\rho_{nn}dE_{n}\right),
\end{equation}
where $E_{n}$ and $\rho_{nn}$ are the eigenenergy and the occupation
probability of the $n$th eigenstate, respectively. The work done
$\bar{d}W$ and the heat exchange $\bar{d}Q$ during an infinitesimal
thermodynamic process are often identified as \citep{key-9,key-10}
\begin{equation}
\bar{d}W=\sum_{n}\rho_{nn}dE_{n}
\end{equation}
and 
\begin{equation}
\bar{d}Q=\sum_{n}E_{n}d\rho_{nn}.
\end{equation}
These imply that the work has to be done by changing the generalized
coordinates of the system, while the heat transfer between the quantum
system and the heat bath induces the rearrangement of the occupation
probabilities \citep{key-11,key-12}. However, these forms of identification
are not strictly comfortable with rigorous mathematical proof. Questions
inevitably arise when one is faced with a system undergoing different
thermodynamic processes studied commonly. In fact, it may induce doubts
whether the above definitions of the heat and work are applicable
for any process. 

Based on the dissipative master equation, the heat and work relevant
to the open quantum system with time-dependent Hamiltonian were introduced
in the pioneering research of Alicki \citep{key-13,key-14}. Kosloff
et al. implemented Alicki\textquoteright s formulas and systematically
studied quantum heat engine cycles working with harmonic oscillators
and spins \citep{key-15,key-16,key-17}. Here, writing the heat and
work in term of the systemic instantaneous orthonormal basis, we show
that quantum coherence characterized by the off-diagonal elements
of the density matrix stimulates additional energy changes in thermodynamic
processes. The quantum effects make the heat and work of quantum isochoric,
quantum isothermal, and thermodynamic adiabatic processes different
from one another. The exactly-solvable dynamics of high-spin precession
will be used to prove that quanutm coherence guarantees the thermodynamic
adiabatic evolution of time-dependent system.

\section{Expressions of heat and work in quantum thermodynamic processes }

For a general quantum system, an external driving field gives rise
to a time-dependent Hamiltonian $\hat{H}(t)$. According to the microscopic
description of the first law of thermodynamics, the change in the
internal energy of the system can be split into two separate parts,
i.e., 
\begin{equation}
\dot{U}=Tr\left(\dot{\hat{\rho}}\hat{H}\right)+Tr\left(\hat{\rho}\dot{\hat{H}}\right),
\end{equation}
where the dot denotes the time derivative and $\hat{\rho}$ is the
density operator corresponding to the ensemble \citep{key-13,key-18,key-19}. 

Let $\left\{ \left|m\left(t\right)\right\rangle ,m=1,2,\cdots\right\} $
be a complete instantaneous orthonormal basis of $\hat{H}(t)$, we
can write the Hamiltonian and the density operator in a matrix form,
that is
\begin{equation}
\hat{H}(t)\text{=}\sum_{m}E_{m}\left(t\right)\left|m\left(t\right)\right\rangle \left\langle m\left(t\right)\right|
\end{equation}
 and
\begin{equation}
\hat{\rho}\text{\ensuremath{\left(t\right)}=}\sum_{nm}\rho_{nm}\left(t\right)\left|n\left(t\right)\right\rangle \left\langle m\left(t\right)\right|,
\end{equation}
where $E_{m}\left(t\right)$ is the eigenvalue of the state $\left|m\left(t\right)\right\rangle $
at any particular instant and $\rho_{nm}\left(t\right)=\left\langle n\left(t\right)\right|\hat{\rho}\left|m\left(t\right)\right\rangle $
represents the density matrix element. For the sake of simplicity,
time $t$ is omitted in the notation. Note that the off-diagonal element
$\rho_{nm}\left(n\neq m\right)$ exists, meaning that $\hat{\rho}$
and $\hat{H}$ do not have a common orthonormal basis. With the help
of Eqs. (5) and (6), we have $Tr\left(\dot{\hat{\rho}}\hat{H}\right)=\sum_{n}\dot{\rho}_{nn}E_{n}-\sum_{n\neq m}\rho_{nm}\left\langle m\right|\frac{\partial\hat{H}}{\partial t}\left|n\right\rangle $,
and $Tr\left(\hat{\rho}\dot{\hat{H}}\right)=\sum_{n}\rho_{nn}\dot{E}_{n}+\sum_{n\neq m}\rho_{nm}\left\langle m\right|\frac{\partial\hat{H}}{\partial t}\left|n\right\rangle $
(Appendix A). The rate of change of the internal energy becomes

\begin{align}
\dot{U} & =\sum_{n}\dot{\rho}_{nn}E_{n}-\sum_{n\neq m}\rho_{nm}\left\langle m\right|\frac{\partial\hat{H}}{\partial t}\left|n\right\rangle \nonumber \\
 & +\sum_{n}\rho_{nn}\dot{E}_{n}+\sum_{n\neq m}\rho_{nm}\left\langle m\right|\frac{\partial\hat{H}}{\partial t}\left|n\right\rangle ,
\end{align}
which are classified into four different categories. The second and
fourth terms representing the quantum coherence have the same magnitude
but different signs. They would cancel each other resulting in the
consistency between Eq. (7) and Eq. (1). However, we find that the
connection between the quantum coherence and the infinitesimal increments
of the heat and work exists. For a closed system subjected to a time-dependent
force, the unawareness of the quantum coherence may violate the first
law of thermodynamics. 

In thermodynamics, an adiabatic process in a closed system occurs
when the transfer of heat and matter between the thermodynamic system
and its surrounding is avoided. The evolution of the density operator
during the adiabatic process is unitary. Based on the Liouville-von
Neumann equation \citep{key-20}
\begin{equation}
\dot{\hat{\rho}}=-\frac{i}{\hbar}\left[\hat{H},\hat{\rho}\right]
\end{equation}
and the time derivative of the density matrix formula, we have (see
Appendix B)
\begin{equation}
\sum_{n}\dot{\rho}_{nn}E_{n}=\sum_{n\neq m}\rho_{nm}\left\langle m\right|\frac{\partial\hat{H}}{\partial t}\left|n\right\rangle ,
\end{equation}
which indicates that $Tr\left(\dot{\hat{\rho}}\hat{H}\right)=0$.
According to the first law of thermodynamics, the internal energy
in an adiabatic process is transferred only as work, and the rate
of work performed in this process is given by

\begin{equation}
\dot{W}=\sum_{n}\rho_{nn}\dot{E}_{n}+\sum_{n\neq m}\rho_{nm}\left\langle m\right|\frac{\partial\hat{H}}{\partial t}\left|n\right\rangle .
\end{equation}

The rate of the heat transfer should have the following form 

\begin{equation}
\dot{Q}=\sum_{n}\dot{\rho}_{nn}E_{n}-\sum_{n\neq m}\rho_{nm}\left\langle m\right|\frac{\partial\hat{H}}{\partial t}\left|n\right\rangle \text{.}
\end{equation}
Only when the quantum coherence, represented by $\sum_{n\neq m}\rho_{nm}\left\langle m\right|\frac{\partial\hat{H}}{\partial t}\left|n\right\rangle $,
is considered, the absence of the heat loss to the surroundings in
the adiabatic process is guaranteed. The above description of the
heat and work are compatible with Alicki and Kieu's definitions and
can be generalized to open quantum system dynamics \citep{key-13,key-21}. 

For the combined system bath scenario, the quantum master equation
is described explicitly as $\dot{\hat{\rho}}=-\frac{i}{\hbar}\left[\hat{H},\hat{\rho}\right]+\mathcal{L_{\mathrm{D}}\left(\hat{\rho}\right)}$,
where $\mathcal{L}_{\mathrm{D}}\left(\hat{\rho}\right)$ is the dissipative
superoperator responsible for the interaction of a quantum system
with its environment. Straightforwardly, we obtain $Tr\left(\dot{\hat{\rho}}\hat{H}\right)=Tr\left(\mathcal{L_{\mathrm{D}}\left(\hat{\rho}\right)}\hat{H}\right)$,
which demonstrates that all dissipative parts due to the heat exchange
are contained in Eq. (11). Equations (10) and (11) give the general
definitions of the heat and work in quantum thermodynamic processes.
The first part of Eq. (10) indicates that the work done on or by a
system can be obtained through the redistribution of the energy eigenvalues
$E_{n}$. The first term in Eq. (11), on the other hand, shows that
the heat transfer is related to a change in the occupation probabilities
$\rho_{nn}$. The second terms in Eqs. (10) and (11) imply that both
the heat transfer and the work done in a microscopic process are closely
related to the quantum coherence. Considering other specific types
of thermodynamic processes, we will find that the heat transfer rate
and the work flux may have different characteristics. 

When the external field is fixed and the quantum system is put into
contacting with a thermal bath at a certain temperature, an isochoric
evolution can be carried out. Since $\frac{\partial\hat{H}}{\partial t}=0$
and the eigenvalues of the Hamiltonian operator $E_{n}$ remain constant
throughout the isochoric process, no external work is performed $\left(\dot{W}=0\right)$,
leading to the sole change in the internal energy due to the heat
exchange. The heat transfer rate between the system and the thermal
bath under this condition can be calculated as $\dot{Q}=\sum_{n}\dot{\rho}_{nn}E_{n}$,
which is simply depending on a change in the population of the microstates.

The quantum isothermal processes typically occur when a system is
kept in contact with a thermal bath. The system is capable of performing
positive work to the outside, and meanwhile absorbs heat from the
bath. Both the eigenvalues $E_{n}$ and the occupation probabilities
$\rho_{nn}$ need to be changed simultaneously. This operation will
occur slowly enough to allow the system to remain in equilibrium with
the thermal bath at every instant. The density operator at thermal
equilibrium is characterized by thermally distributed populations
in the quantum states $\hat{\rho}\left(t\right)=1/Z\sum_{m}\exp\left[-E_{m}\left(t\right)/\left(k_{B}T\right)\right]\left|m\left(t\right)\right\rangle \left\langle m\left(t\right)\right|$
with $Z=\sum_{m}\exp\left[-E_{m}\left(t\right)/\left(k_{B}T\right)\right]$
being the canonical partition function. Since the system approaches
thermal equilibrium without a typical relaxation time due to the bath-system
interactions, quantum coherence vanishes, i.e., $\rho_{nm}=0\left(n\neq m\right)$.
The heat transfer rate and the work flux in the quantum isothermal
processes can be expressed as $\dot{Q}=\sum_{n}\dot{\rho}_{nn}E_{n}$
and $\dot{W}=\sum_{n}\rho_{nn}\dot{E}_{n}$, respectively. Because
the adiabatic process does not require the second terms in Eqs. (10)
and (11) to be a zero value, the general expressions of the rates
of the work performed and the heat transfer are different from the
counterparts in the isothermal and isochoric processes. Examples of
illustrating the valid arguments of the above discussion will be given
in the following sections. 

\begin{figure}
\includegraphics[scale=0.3]{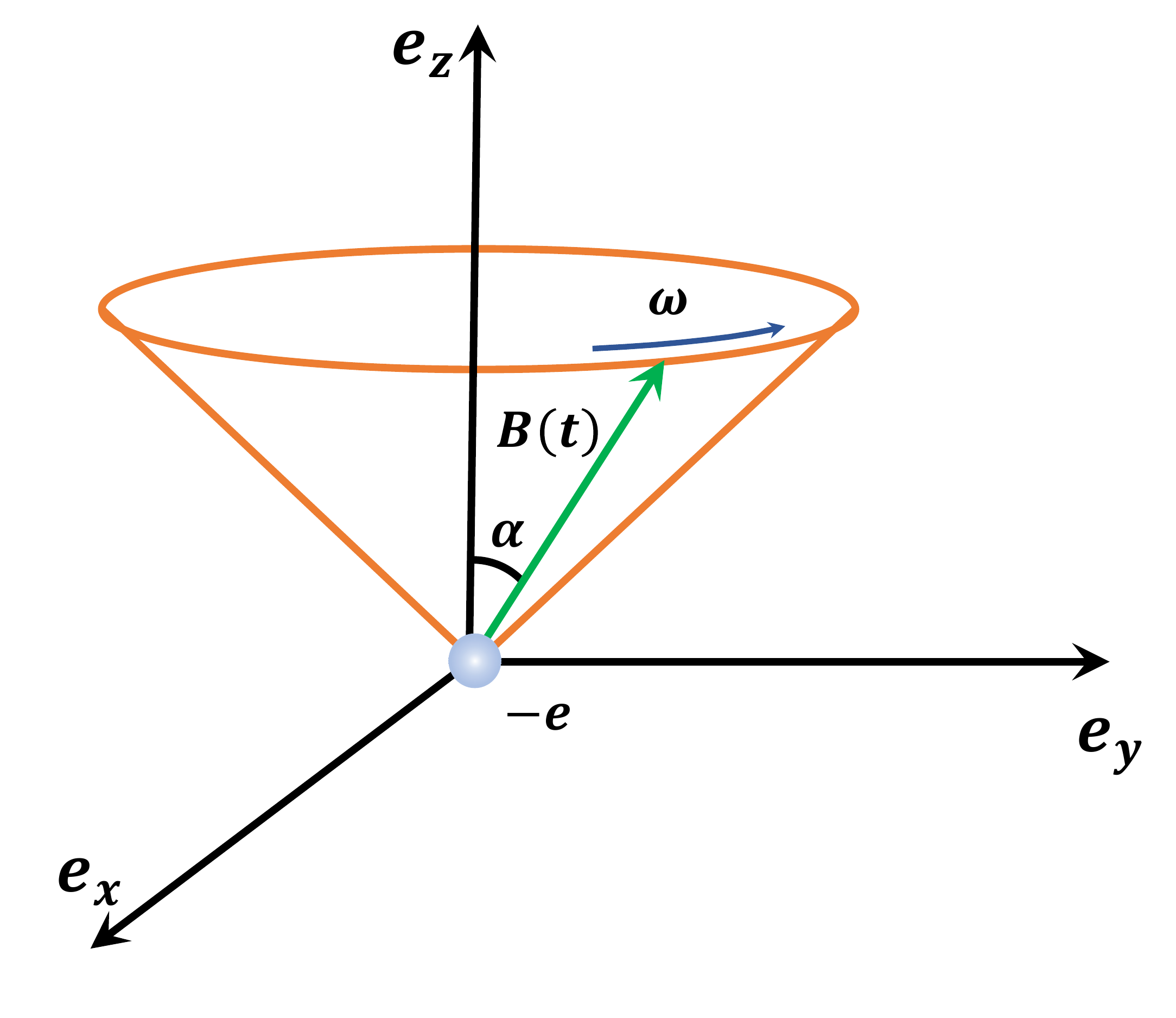}

\caption{ An electron in the presence of a magnetic field whose magnitude $B_{0}$
is a constant, but whose direction rides around at an angular velocity
$\omega$ on the tip of a cone of opening angle $\alpha$. }
\end{figure}

\section{The adiabatic evolution of a two-level model in a magnetic field }

It is instructive to look at an electron with charge $-e$ and mass
$m$ located at the origin of three-dimensional space (Fig.1). The
dipole moment of the electron is proportional to its gyromagnetic
ratio $\gamma_{e}=-e/m$ and spin angular momentum $\hat{\boldsymbol{S}}$,
i.e., $\mathbf{\hat{\mathbf{\mathbf{\boldsymbol{\mu}}}}}=\gamma_{e}\hat{\boldsymbol{S}}$.
When a magnetic field $\boldsymbol{B}\left(t\right)=B_{0}\left[\sin\alpha\cos\left(\omega t\right)\boldsymbol{e}_{x}+\sin\alpha\sin\left(\omega t\right)\boldsymbol{e}_{y}+\cos\alpha\boldsymbol{e}_{z}\right]$
is applied, the electric dipole of the electron interacts with the
field and the time-dependent Hamiltonian of the system is \citep{key-22}

\begin{align}
\hat{H_{e}}\left(t\right) & =-\hat{\mathbf{\mathbf{\boldsymbol{\mu}}}}\cdot\boldsymbol{B}\left(t\right)\\
 & =\frac{\hbar\omega_{1}}{2}[\sin\alpha\cos\left(\omega t\right)\hat{\sigma}_{x}+\sin\alpha\sin\left(\omega t\right)\hat{\sigma}_{y}+\cos\alpha\hat{\sigma}_{z}],\nonumber 
\end{align}
where $\omega_{1}=eB_{0}/m$; $\hat{\sigma}_{i}$ ($i=x$, $y$, and
$z$) are the usual Pauli spin matrices; and $\hbar$ equals the Planck
constant divided by $2\pi$. 

The wave function for the two-state system is a linear combination
of the normalized eigenvectors $\left|\chi_{+}\left(t\right)\right\rangle $
and $\left|\chi_{-}\left(t\right)\right\rangle $, i.e.,

\begin{align}
\left|\chi\left(t\right)\right\rangle  & =\left[\cos\left(\frac{\lambda t}{2}\right)-i\frac{\omega_{1}-\omega\cos\alpha}{\lambda}\sin\left(\frac{\lambda t}{2}\right)\right]e^{-i\omega t/2}\nonumber \\
 & \times\left|\chi_{+}\left(t\right)\right\rangle +i\left[\frac{\omega}{\lambda}\sin\alpha\sin\left(\frac{\lambda t}{2}\right)\right]e^{i\omega t/2}\left|\chi_{-}\left(t\right)\right\rangle ,
\end{align}
where $\lambda=\sqrt{\omega^{2}+\omega_{1}^{2}-2\omega\omega_{1}\cos\alpha}$
(see Appendix C). The density matrix operator can be written as $\hat{\rho}=\left|\chi\left(t\right)\right\rangle \left\langle \chi\left(t\right)\right|$.
Taking the matrix elements, we get $\rho_{++}=\left\langle \chi_{+}\left(t\right)\right|\hat{\rho}\left|\chi_{+}\left(t\right)\right\rangle $,
$\rho_{--}=\left\langle \chi_{-}\left(t\right)\right|\hat{\rho}\left|\chi_{-}\left(t\right)\right\rangle $,
$\rho_{+-}=\left\langle \chi_{+}\left(t\right)\right|\hat{\rho}\left|\chi_{-}\left(t\right)\right\rangle $,
and $\rho_{-+}=\left\langle \chi_{-}\left(t\right)\right|\hat{\rho}\left|\chi_{+}\left(t\right)\right\rangle $. 

It is obvious that $\rho_{++}$ and $\rho_{--}$ are the probabilities
of being in the spin up and spin down states along $\boldsymbol{B}\left(t\right)$
and are, respectively, given by 
\begin{equation}
\rho_{++}=\cos^{2}\left(\frac{\lambda t}{2}\right)+\frac{\left(\omega_{1}-\omega\cos\alpha\right)^{2}}{\lambda^{2}}\sin^{2}\left(\frac{\lambda t}{2}\right),
\end{equation}
and

\begin{equation}
\rho_{--}=\left[\frac{\omega}{\lambda}\sin\alpha\sin\left(\frac{\lambda t}{2}\right)\right]^{2}.
\end{equation}

Most existing literatures studied the thermodynamic properties of
the quantum systems with time-independent Hamiltonian. Thus, the heat
exchange rate and the work flux in a thermodynamic process can be
simplified as $\dot{Q}=\sum_{n}\dot{\rho}_{nn}E_{n}$ and $\dot{W}=\sum_{n}\rho_{nn}\dot{E}_{n}$.
These indicates that only the population transfer results in the microscopic
realization of the heat exchange, while the work merely depends on
the energy change generated by the external field. However, for the
electron-spin system driven by a rotating magnetic field, the eigenvalues
corresponding to the instantaneous eigenstates $\left|\chi_{+}\left(t\right)\right\rangle $
and $\left|\chi_{-}\left(t\right)\right\rangle $ are time independent
(see Appendix C). If the work flux in the adiabatic evolution process
remains being computed by $\dot{W}=\sum_{n}\rho_{nn}\dot{E}_{n}$,
we have $\dot{W}=0$. It means that no work can be done by the magnetic
field $\boldsymbol{B}\left(t\right)$, which is an apparent contradiction
in physical reality. In addition, according to the time derivative
of $\rho_{++}$ and $\rho_{--}$ from Eqs. (14) and (15), the heat
exchange rate $\dot{Q}=\dot{\rho}_{++}E_{+}+\dot{\rho}_{--}E_{-}=-\frac{\hbar\omega_{1}}{2\lambda}\omega^{2}\sin^{2}\alpha\sin\left(\lambda t\right)$.
As $\dot{Q}$ is a non-zero value, it is a paradox that there exists
heat transfer between the system and the environment in the thermodynamic
adiabatic process. 

In the present studies, we emphasize that the quantum coherence effects
are necessary for reclaiming the validity of the first law of thermodynamics
when a quantum system with time-dependent Hamiltonian is considered.
Making use of Eq. (13) and taking the off-diagonal elements of the
density operator, one readily get 
\begin{align}
\sum_{n\neq m}\rho_{nm}\left\langle m\right|\frac{\partial\hat{H_{e}}}{\partial t}\left|n\right\rangle  & =\rho_{-+}\left\langle \chi_{+}\right|\frac{\partial\hat{H_{e}}}{\partial t}\left|\chi_{-}\right\rangle \nonumber \\
 & +\rho_{+-}\left\langle \chi_{-}\right|\frac{\partial\hat{H_{e}}}{\partial t}\left|\chi_{+}\right\rangle \nonumber \\
 & =-\frac{\hbar\omega_{1}}{2\lambda}\omega^{2}\sin^{2}\alpha\sin\left(\lambda t\right).
\end{align}
It is observed that $\sum_{n}\dot{\rho}_{nn}E_{n}=\sum_{n\neq m}\rho_{nm}\left\langle m\right|\frac{\partial\hat{H_{e}}}{\partial t}\left|n\right\rangle $.
According to this and Eq. (11), we ensure that the heat transfer rate
$\dot{Q}$ of the two-state system in the adiabatic evolution process
equals zero and the corresponding power generated by the external
field $\dot{W}=-\frac{\hbar\omega_{1}}{2\lambda}\omega^{2}\sin^{2}\alpha\sin\left(\lambda t\right)$. 

\section{Test of the thermodynamic adiabatic process via the exactly-solvable
dynamic of high-spin precession }

In this section, we consider a neutral particle with a magnetic moment
and arbitrary spin $j$ in a harmonically-changing external magnetic
field $\boldsymbol{B}\left(t\right)=B_{0}\left[\begin{array}{ccc}
\sin\theta\cos\left(\omega t\right), & \sin\theta\sin\left(\omega t\right), & \cos\theta\end{array}\right]$. The magnetic field rotates around the z-axis with frequency $\omega$
and is inclined at a constant angle $\theta$. The systemic Hamiltonian
is time dependent and is given by

\begin{align}
\hat{H_{J}}\left(t\right) & =\gamma\boldsymbol{B}\left(t\right)\cdot\hat{\boldsymbol{J}}\\
 & =\gamma B_{0}[\hat{J}_{x}\sin\theta\cos\left(\omega t\right)+\hat{J}_{y}\sin\theta\sin\left(\omega t\right)+\hat{J}_{z}\cos\theta],\nonumber 
\end{align}
where $\gamma$ is the coupling parameter and $\hat{\boldsymbol{J}}$
is the total angular momentum vector. The operators $\hat{J}_{x}$,
$\hat{J}_{y}$, and $\hat{J}_{z}$ represent three Cartesian components
of the angular momentum. Learning the thermodynamic behavior of high-spin
precession faces a demand with the exact solution for the wave function.
The Schrödinger equation with the time-dependent Hamiltonian is usually
difficult to deal with. However, it can be reduced to a Schrödinger-like
equation with a time-independent effective Hamiltonian by invoking
quantum rotation transformation in angular momentum theory \citep{key-23}. 

If the system is placed initially in an instantaneous eigenstate of
$\hat{H_{J}}\left(0\right)$, i.e., 
\begin{equation}
\left|\psi\left(0\right)\right\rangle \text{=}\hat{R}_{z}\left(0\right)\hat{R}_{y}\left(\varphi\right)\left|j,M\right\rangle \text{=}\sum_{m'}d_{m'M}^{j}\left(\beta\right)\left|j,m'\left(\varphi\right)\right\rangle ,
\end{equation}
the exact solution of the time-dependent wavefunction becomes 
\begin{equation}
\left|\psi\left(t\right)\right\rangle \text{=}\sum_{mm'}d_{m'M}^{j}\left(\beta\right)d_{mm'}^{j}\left(\varphi\right)e^{-im'\omega_{0}t}e^{-i\hat{J}_{z}\omega t/\hbar}\left|j,m\right\rangle ,
\end{equation}
where $\lambda_{0}=\frac{\omega}{\gamma B_{0}}$; $\omega_{0}^{2}=\left(\gamma B_{0}\right)^{2}\left(1-2\lambda_{0}\cos\theta+\lambda_{0}^{2}\right)$;
and $\beta=\theta-\varphi$. $\hat{R}_{z}\left(\omega t\right)=e^{-i\hat{J}_{z}\omega t/\hbar}$
and $\hat{R}_{y}\left(\varphi\right)=e^{-i\hat{J}_{y}\varphi/\hbar}$
are the rotation operators. $\left|j,m\left(\varphi\right)\right\rangle =\hat{R}_{y}\left(\varphi\right)\left|j,m\right\rangle $
$\left(m=j,j-1,\cdots,-j\right)$ are the rotations of the standard
angular momentum basis $\left|j,m\right\rangle $. $d_{m'm}^{j}\left(\varphi\right)=\left\langle j,m'\right|e^{-i\hat{J}_{y}\varphi/\hbar}\left|j,m\right\rangle $
represents an element of Wigner\textquoteright s d-matrix. Details
of the algorithm are given in Appendix D. The wavefunction $\left|\psi\left(t\right)\right\rangle $
is a linear combination of the eigenstates $\left|\psi_{m}\right\rangle $.
As the density operator $\hat{\rho}=\left|\psi\left(t\right)\right\rangle \left\langle \psi\left(t\right)\right|$,
we can explicitly carry out its matrix elements 
\begin{align}
\rho_{nl}\left(t\right) & =\left\langle \psi_{n}\right|\hat{\rho}\left|\psi_{l}\right\rangle \nonumber \\
 & =\sum_{mm'}e^{-i\left(m-m'\right)\omega_{0}t}d_{mM}^{j}\left(\beta\right)d_{m'M}^{j}\left(\beta\right)d_{mn}^{j}\left(\beta\right)\nonumber \\
 & \times d_{m'l}^{j}\left(\beta\right).
\end{align}

Similar to the previous model, the dynamic evolution of high-spin
precession can also be visualized as an adiabatic process in which
no heat is gained or lost by the system. Accordingly, using Eq. (11),
we will show that the high-spin system is hardly thermally isolated
unless the quantum coherence is considered. From Eqs. (11) and (20),
we immediately have

\begin{align}
\sum_{n}\dot{\rho}_{nn}E_{n} & =-i\omega_{0}\gamma B_{0}\hbar\sum_{nmm'}\left(m-m'\right)ne^{-i\left(m-m'\right)\omega_{0}t}\nonumber \\
 & \times d_{mM}^{j}\left(\beta\right)d_{m'M}^{j}\left(\beta\right)d_{mn}^{j}\left(\beta\right)d_{m'n}^{j}\left(\beta\right).
\end{align}

Knowing that 
\begin{align}
\left\langle \psi_{m}\right|\frac{\partial\hat{H_{J}}}{\partial t}\left|\psi_{n}\right\rangle  & =\frac{\omega\gamma B_{0}\hbar\sin\theta}{2i}[\sqrt{\left(j-n\right)\left(j+n+1\right)}\delta_{m,n+1}\nonumber \\
 & -\sqrt{\left(j+n\right)\left(j-n+1\right)}\delta_{m,n-1}]
\end{align}
and the relation $\omega\sin\theta=-\omega_{0}\sin\beta$, we can
write the second term of Eq. (11) as 

\begin{align}
\sum_{nm}\rho_{nm}\left\langle \psi_{m}\right|\frac{\partial\hat{H_{J}}}{\partial t}\left|\psi_{n}\right\rangle =i\omega_{0}\gamma B_{0}\hbar\sum_{nmm'}\frac{\sin\beta}{2}\nonumber \\
e^{-i\left(m-m'\right)\omega_{0}t}d_{mM}^{j}\left(\beta\right)d_{m'M}^{j}\left(\beta\right)d_{mn}^{j}\left(\beta\right)\nonumber \\
\times[d_{m'n+1}^{j}\left(\beta\right)\sqrt{\left(j-n\right)\left(j+n+1\right)}\nonumber \\
-d_{m'n-1}^{j}\left(\beta\right)\sqrt{\left(j+n\right)\left(j-n+1\right)}].
\end{align}

In the spin $1/2$ case, the matrix representation of Wigner d-function
$d^{1/2}\left(\beta\right)=\left(\begin{array}{cc}
\cos\left(\beta/2\right) & -\sin\left(\beta/2\right)\\
\sin\left(\beta/2\right) & \cos\left(\beta/2\right)
\end{array}\right)$. If we substitute $\omega_{0}\rightarrow\lambda$, $\gamma B_{0}\rightarrow\omega_{1}$,
and $\varphi\rightarrow\alpha$, it is not difficult to work out that
$\sum_{n}\dot{\rho}_{nn}E_{n}=\sum_{n\neq m}\rho_{nm}\left\langle \psi_{m}\right|\frac{\partial\hat{H_{J}}}{\partial t}\left|\psi_{n}\right\rangle =-\frac{\hbar\omega_{1}}{2\lambda}\omega^{2}\sin^{2}\alpha\sin\left(\lambda t\right)$.
This formula is completely analogy to the result obtained by the two-level
model moving in an adiabatically rotating magnetic field. As an interesting
application of the exact explicit solutions of the Hamiltonian in
Eq. (17), we will show that the spin-precession processes are thermodynamic
adiabatic regardless of the spin quantum numbers of any particles.
The proof is straightforward, which can be done by substituting Eqs.
(21) and (23) into Eq. (11). 

When $m=m'$, we have $\sum_{m}\rho_{mm}\left\langle \psi_{m}\right|\frac{\partial\hat{H_{J}}}{\partial t}\left|\psi_{m}\right\rangle =0$.
Only the terms with $m\neq m'$ need to be considered in the computation.
In the quantum theory of angular momentum, the recursion relation
for Wigner\textquoteright s d-matrix implies that \citep{key-24,key-25,key-26}
\begin{align}
\frac{-m+n\cos\beta}{\sin\beta}d_{mn}^{j}\left(\beta\right) & =\frac{1}{2}[d_{mn+1}^{j}\left(\beta\right)\sqrt{\left(j-n\right)\left(j+n+1\right)}\nonumber \\
 & +d_{mn-1}^{j}\left(\beta\right)\sqrt{\left(j+n\right)\left(j-n+1\right)}].
\end{align}

Combining the relation between $d_{mn}^{j}\left(\beta\right)$ and
$d_{mn\pm1}^{j}\left(\beta\right)$ and the usual index transformation,
one can derive the following invariant sum which gives the coupling
rules relating to the direct product of two rotation matrices 

\begin{align}
\sum_{n}\left(m-m'\right)nd_{mn}^{j}\left(\beta\right)d_{m'n}^{j}\left(\beta\right)=\frac{\sin\beta}{2}\sum_{n}d_{mn}^{j}\left(\beta\right)\nonumber \\
\times[d_{m'n-1}^{j}\left(\beta\right)\sqrt{\left(j+n\right)\left(j-n+1\right)}\nonumber \\
-d_{m'n+1}^{j}\left(\beta\right)\sqrt{\left(j-n\right)\left(j+n+1\right)}]\text{.}
\end{align}
The detail calculations are given in the Appendix E. Applying the
invariant sum to Eqs. (21) and (23), we can verify that $\sum_{n}\dot{\rho}_{nn}E_{n}=\sum_{n\neq m}\rho_{nm}\left\langle \psi_{m}\right|\frac{\partial\hat{H}_{J}}{\partial t}\left|\psi_{n}\right\rangle $
for an instantaneous state. As the heat exchange rate $\dot{Q}=0$,
the spin system exchanges no mass or heat energy with its environment.
The change in its internal energy is merely due to the work done by
the external magnetic field. Once again, the analysis demonstrates
that the unitary evolution of a closed system with time-dependent
Hamiltonian is equivalent to a thermodynamic adiabatic process when
the quantum coherence is taken into account. 

If a system starts in an eigenstate of the initial Hamiltonian, the
quantum adiabatic theorem states that the system will remain in the
corresponding instantaneous eigenstate of the final Hamiltonian when
a given perturbation acting on it is slowly enough \citep{key-27,key-28}.
The adiabatic approximation holds when the time derivative of Hamiltonian
is extremely small and the dimensionless adiabatic parameter $\tau=\left|\hbar\left\langle m\right|\frac{\partial\hat{H}}{\partial t}\left|n\right\rangle /\left(E_{n}-E_{m}\right)^{2}\right|\ll1\left(n\neq m\right)$.
However, the above analysis shows that the thermodynamic adiabatic
processes do not require the quantum adiabatic approximation to be
satisfied. Quantum adiabatic process certainly results in a thermodynamic
adiabatic process, but not all thermodynamic adiabatic processes are
due to the quantum-mechanical adiabatic processes. 

\section{Conclusions}

In summary, we found that the heat and work in microscopic processes
are closely related to the transition between different quantum states.
The energy exchanges during the quantum isochoric and isothermal processes
are simply depending on the change in the eigenenergies or the probability
distributions. However, for a closed system with a time-dependent
driving, the unitary evolution is equivalent to a thermodynamic adiabatic
process only when the quantum coherence is taken into account. Under
this consideration, one can ensure that no heat is lost to or gained
from the surroundings in the case of quantum-spin precession. The
microscopic expressions for thermodynamic quantities are applicable
to both the thermal equilibrium case and the nonequilibrium case. 
\begin{acknowledgments}
This work has been supported by the National Natural Science Foundation
of China (Grants No. 11421063 and No. 11534002), the National 973
program (Grants No. 2012CB922104 and No. 2014CB921403), and the Postdoctoral
Science Foundation of China (Grant No. 2015M580964). 
\end{acknowledgments}

\appendix

\section{EXPRESSING $Tr(\dot{\hat{\rho}}\hat{H})$ AND $Tr(\hat{\rho}\dot{\hat{H}})$
IN TERMS OF THE INSTANTANEOUS ORTHONORMAL BASIS OF $\hat{H}$.}

From Eq. (6), the time derivative of the density operator is given
by
\begin{equation}
\dot{\hat{\rho}}=\sum_{nm}\left(\dot{\rho}_{nm}\left|n\right\rangle \left\langle m\right|+\rho_{nm}\left|\dot{n}\right\rangle \left\langle m\right|+\rho_{nm}\left|n\right\rangle \left\langle \dot{m}\right|\right).
\end{equation}

Using Eq. (A1), we obtain the trace of $\dot{\hat{\rho}}\hat{H}$
as

\begin{align}
Tr\left(\dot{\hat{\rho}}\hat{H}\right) & =\sum_{m'}\left\langle m'\right|\dot{\hat{\rho}}E_{m'}\left|m'\right\rangle \nonumber \\
 & =\sum_{m'}\sum_{nm}E_{m'}\dot{\rho}_{nm}\left\langle m'\right.\left|n\right\rangle \left\langle m\right|\left.m'\right\rangle \nonumber \\
 & +\sum_{m'}\sum_{nm}E_{m'}\rho_{nm}\left\langle m'\right.\left|\dot{n}\right\rangle \left\langle m\right|\left.m'\right\rangle \nonumber \\
 & +\sum_{m'}\sum_{nm}E_{m'}\rho_{nm}\left\langle m'\right.\left|n\right\rangle \left\langle \dot{m}\right|\left.m'\right\rangle \nonumber \\
 & =\sum_{n}E_{n}\dot{\rho}_{nn}+\sum_{nm}\text{\ensuremath{E_{m}\rho_{nm}\left\langle m\right.\left|\dot{n}\right\rangle }}\nonumber \\
 & +\sum_{nm}E_{n}\rho_{nm}\left\langle \dot{m}\right|\left.n\right\rangle )\nonumber \\
 & =\sum_{n}\dot{\rho}_{nn}E_{n}-\sum_{n\neq m}\rho_{nm}\left\langle m\right|\frac{\partial\hat{H}}{\partial t}\left|n\right\rangle .
\end{align}
In the last step, the relation \citep{key-22}
\begin{equation}
\left\langle m|\dot{n}\right\rangle =-\left\langle \dot{m}|n\right\rangle =\left\langle m\right|\frac{\partial\hat{H}}{\partial t}\left|n\right\rangle /\left(E_{n}-E_{m}\right)\left(n\neq m\right)
\end{equation}
has been applied. 

Next, we consider the trace formula $Tr\left(\hat{\rho}\dot{\hat{H}}\right)$.
The time derivative of the systemic Hamiltonian in Eq. (5) reads
\begin{equation}
\dot{\hat{H}}\text{=}\sum_{\alpha}\dot{E}_{\alpha}\left|\alpha\right\rangle \left\langle \alpha\right|+\sum_{\alpha}E_{\alpha}\left|\dot{\alpha}\right\rangle \left\langle \alpha\right|+\sum_{\alpha}E_{\alpha}\left|\alpha\right\rangle \left\langle \dot{\alpha}\right|\text{.}
\end{equation}
With the help of Eqs. (A3) and (A4), it is reasonable to expect that

\begin{align}
Tr\left(\hat{\rho}\dot{\hat{H}}\right) & =\sum_{nm}\rho_{nm}\left\langle m\right|\dot{\hat{H}}\left|n\right\rangle \nonumber \\
 & =\sum_{nm}\rho_{nm}\sum_{\alpha}\dot{E}_{\alpha}\left\langle m\right.\left|\alpha\right\rangle \left\langle \alpha\right|\left.n\right\rangle \nonumber \\
 & +\sum_{nm}\rho_{nm}\sum_{\alpha}E_{\alpha}\left\langle m\right.\left|\dot{\alpha}\right\rangle \left\langle \alpha\right|\left.n\right\rangle \nonumber \\
 & +\sum_{nm}\rho_{nm}\sum_{\alpha}E_{\alpha}\left\langle m\right.\left|\alpha\right\rangle \left\langle \dot{\alpha}\right|\left.n\right\rangle \nonumber \\
 & =\sum_{n}\rho_{nn}\dot{E}_{n}+\sum_{nm}\rho_{nm}E_{n}\left\langle m\right.\left|\dot{n}\right\rangle \nonumber \\
 & +\sum_{nm}\rho_{nm}E_{m}\left\langle \dot{m}\right|\left.n\right\rangle \nonumber \\
 & =\sum_{n}\rho_{nn}\dot{E}_{n}+\sum_{n\neq m}\rho_{nm}\left\langle m\right|\frac{\partial\hat{H}}{\partial t}\left|n\right\rangle .
\end{align}

\section{TIME DEPENDENCE OF THE DENSITY MATRIX IN AN ADIABATIC PROCESS.}

By invoking Eqs. (5) and (6) in the main text, the Poisson bracket
$\left[\hat{H},\hat{\rho}\right]$ is expanded as follows, 

\begin{align}
\left[\hat{H},\hat{\rho}\right] & =\hat{H}\hat{\rho}-\hat{\rho}\hat{H}\nonumber \\
 & =\sum_{m'}E_{m'}\left|m'\right\rangle \left\langle m'\right|\sum_{nm}\rho_{nm}\left|n\right\rangle \left\langle m\right|\nonumber \\
 & -\sum_{nm}\rho_{nm}\left|n\right\rangle \left\langle m\right|\sum_{m'}E_{m'}\left|m'\right\rangle \left\langle m'\right|\nonumber \\
 & =\sum_{nm}\left(E_{n}-E_{m}\right)\rho_{nm}\left|n\right\rangle \left\langle m\right|\text{.}
\end{align}
Substituting Eq. (B1) into the Liouville-von Neumann equation, we
obtain 

\begin{align}
\sum_{nm}\left(\dot{\rho}_{nm}\left|n\right\rangle \left\langle m\right|+\rho_{nm}\left|\dot{n}\right\rangle \left\langle m\right|+\rho_{nm}\left|n\right\rangle \left\langle \dot{m}\right|\right)\nonumber \\
=-\frac{i}{\hbar}\sum_{nm}\left(E_{n}-E_{m}\right)\rho_{nm}\left|n\right\rangle \left\langle m\right|.
\end{align}

Multiplying both sides of Eq. (B2) by a bra $\left\langle k\right|$
on the left and a ket $\left|l\right\rangle $ on the right yields

\begin{equation}
\dot{\rho}_{kl}=-\frac{i}{\hbar}\left(E_{k}-E_{l}\right)\rho_{kl}-\sum_{m}\left(\rho_{ml}\left\langle k\right.\left|\dot{m}\right\rangle +\rho_{km}\left\langle \dot{m}\right|\left.l\right\rangle \right).
\end{equation}
When $k=l=n$ and considering $\left\langle m|\dot{n}\right\rangle =-\left\langle \dot{m}|n\right\rangle =\left\langle m\right|\frac{\partial\hat{H}}{\partial t}\left|n\right\rangle /\left(E_{n}-E_{m}\right)\left(n\neq m\right)$,
one readily finds the equality of Eq. (9).

\section{ELECTRON SPIN PRECESSION IN AN ADIABATICALLY ROTATING ELECTRIC FIELD.}

Considering the orthonormal bases $\left|\uparrow\right\rangle =\left(\begin{array}{c}
1\\
0
\end{array}\right)$ and $\left|\downarrow\right\rangle =\left(\begin{array}{c}
0\\
1
\end{array}\right)$, we have the normalized eigenvectors of $\hat{H_{e}}\left(t\right)$
as
\begin{equation}
\left|\chi_{+}\left(t\right)\right\rangle =\cos\frac{\alpha}{2}\left|\uparrow\right\rangle +e^{i\omega t}\sin\frac{\alpha}{2}\left|\downarrow\right\rangle 
\end{equation}
and

\begin{equation}
\left|\chi_{-}\left(t\right)\right\rangle =e^{-i\omega t}\sin\frac{\alpha}{2}\left|\uparrow\right\rangle -\cos\frac{\alpha}{2}\left|\downarrow\right\rangle ,
\end{equation}
which represent the spin up and spin down, respectively, along the
instantaneous direction of $\boldsymbol{B}\left(t\right)$. The corresponding
eigenvalues are 
\begin{equation}
E_{\pm}=\pm\frac{\hbar\omega_{1}}{2}.
\end{equation}

When the electron starts out with spin up along $\boldsymbol{B}\left(t\right)$,
the system is initially in a state of superposition and is given by
$\left|\chi\left(0\right)\right\rangle =\cos\frac{\alpha}{2}\left|\uparrow\right\rangle +\sin\frac{\alpha}{2}\left|\downarrow\right\rangle $.
The Schrödinger equation can be integrated formally to give the time-dependent
wave function $\left|\chi\left(t\right)\right\rangle =\hat{U}\left(t\right)\left|\chi\left(0\right)\right\rangle $,
where the unitary time-evolution operator is solved by $i\hbar\frac{\partial}{\partial t}\hat{U}\left(t\right)=\hat{H_{e}}\left(t\right)\hat{U}\left(t\right)$. 

\section{EXACTLY-SOLVABLE DYNAMICS OF HIGH-SPIN PRECESSION. }

Moving to a rotating frame using $\hat{H}_{R}=\hat{R}_{z}\left(\omega t\right)^{\dagger}\hat{H_{J}}\left(t\right)\hat{R}_{z}\left(\omega t\right)-i\hbar\hat{R}_{z}\left(\omega t\right)^{\dagger}\frac{\partial}{\partial t}\hat{R}_{z}\left(\omega t\right)$
with $\hat{R}_{z}\left(\omega t\right)=e^{-i\hat{J}_{z}\omega t/\hbar}$,
one has 
\begin{equation}
\hat{H}_{R}=\omega_{0}\left(\hat{J}_{x}\sin\varphi+\hat{J}_{z}\cos\varphi\right),
\end{equation}
where $\sin\varphi=\frac{\sin\theta}{\sqrt{1-2\lambda_{0}\cos\theta+\lambda_{0}^{2}}}$,
and $\cos\varphi=\frac{\cos\theta-\lambda_{0}}{\sqrt{1-2\lambda_{0}\cos\theta+\lambda_{0}^{2}}}$.
Obviously, the effective Hamiltonian $\hat{H}_{R}$ contains no explicit
time dependence in the rotating frame. The corresponding rotated wavefunction
is
\begin{equation}
\left|\phi\left(t\right)\right\rangle =\hat{R}_{z}\left(\omega t\right)^{\dagger}\left|\psi\left(t\right)\right\rangle .
\end{equation}

Invoking the transformation $\hat{R}_{y}\left(\varphi\right)=e^{-i\hat{J}_{y}\varphi/\hbar}$,
one can rewrite $\hat{H}_{R}$ as 
\begin{equation}
\hat{H}_{R}=\hat{R}_{y}\left(\varphi\right)\omega_{0}\hat{J}_{z}\hat{R}_{y}\left(\varphi\right)^{\dagger}.
\end{equation}
It\textquoteright s reasonable to expect that $\hat{H}_{R}$\textquoteright s
eigenstates $\left|j,m\left(\varphi\right)\right\rangle =\hat{R}_{y}\left(\varphi\right)\left|j,m\right\rangle $
$\left(m=j,j-1,\cdots,-j\right)$ are the rotations of the standard
angular momentum basis $\left|j,m\right\rangle $. Therefore, when
the initial state of the system is $\left|\psi\left(0\right)\right\rangle $,
the exact wavefunction is straightforwardly given by 
\begin{align}
\left|\psi\left(t\right)\right\rangle \text{=}\sum_{mm'}\left\langle j,m\left(\varphi\right)|\psi\left(0\right)\right\rangle d_{m'm}^{j}\left(\varphi\right)e^{-im\omega_{0}t}\nonumber \\
\times e^{-i\hat{J}_{z}\omega t/\hbar}\left|j,m'\right\rangle .
\end{align}
Starting from $\hat{J}_{z}\left|j,m\right\rangle =m\hbar\left|j,m\right\rangle $
and remembering that $e^{-i\hat{J}_{y}\varphi/\hbar}\hat{J}_{z}e^{i\hat{J}_{y}\varphi/\hbar}=\hat{J}_{z}\cos\varphi+\hat{J}_{x}\sin\varphi$
and $e^{-i\hat{J}_{z}\omega t/\hbar}\hat{J}_{x}e^{i\hat{J}_{z}\omega t/\hbar}=\hat{J}_{x}\cos\omega t+\hat{J}_{y}\sin\omega t$,
we recognize the following relation 
\begin{equation}
\hat{H_{J}}\left(t\right)\hat{R}_{z}\left(\omega t\right)\hat{R}_{y}\left(\varphi\right)\left|j,m\right\rangle =\gamma B_{0}\hat{R}_{z}\left(\omega t\right)\hat{R}_{y}\left(\varphi\right)\hat{J}_{z}\left|j,m\right\rangle .
\end{equation}
It indicates that $\psi_{m}=\hat{R}_{z}\left(\omega t\right)\hat{R}_{y}\left(\varphi\right)\left|j,m\right\rangle $
is the eigenstate of $\hat{H_{J}}\left(t\right)$ and its corresponding
eigenvalue is $m\gamma B_{0}\hbar$. 

\section{PROOF OF THE COUPLING RULES RELATING TO THE DIRECT PRODUCT OF TWO
ROTATION MATRICES.}

In angular momentum theory, one has the relations between $d_{m\left(m'\right)n}^{j}\left(\beta\right)$
and $d_{m\left(m'\right)n\pm1}^{j}\left(\beta\right)$ as \citep{key-24,key-25,key-26}

\begin{align}
\frac{-m+n\cos\beta}{\sin\beta}d_{mn}^{j}\left(\beta\right) & =\frac{1}{2}[d_{mn+1}^{j}\left(\beta\right)\sqrt{\left(j-n\right)\left(j+n+1\right)}\nonumber \\
 & +d_{mn-1}^{j}\left(\beta\right)\sqrt{\left(j+n\right)\left(j-n+1\right)}]\text{,}
\end{align}
and

\begin{align}
\frac{-m'+n\cos\beta}{\sin\beta}d_{m'n}^{j}\left(\beta\right) & =\frac{1}{2}[d_{m'n+1}^{j}\left(\beta\right)\sqrt{\left(j-n\right)\left(j+n+1\right)}\nonumber \\
 & +d_{m'n-1}^{j}\left(\beta\right)\sqrt{\left(j+n\right)\left(j-n+1\right)}].
\end{align}

Multiplying each of them by $d_{m'n}^{j}\left(\beta\right)$ and $d_{mn}^{j}\left(\beta\right)$,
respectively, yields

\begin{align}
\frac{-m+n\cos\beta}{\sin\beta}d_{mn}^{j}\left(\beta\right)d_{m'n}^{j}\left(\beta\right)\nonumber \\
=\frac{1}{2}[d_{mn+1}^{j}\left(\beta\right)d_{m'n}^{j}\left(\beta\right)\sqrt{\left(j-n\right)\left(j+n+1\right)}\nonumber \\
+d_{mn-1}^{j}\left(\beta\right)d_{m'n}^{j}\left(\beta\right)\sqrt{\left(j+n\right)\left(j-n+1\right)}] & ,
\end{align}
and

\begin{align}
\frac{-m'+n\cos\beta}{\sin\beta}d_{m'n}^{j}\left(\beta\right)d_{mn}^{j}\left(\beta\right)\nonumber \\
=\frac{1}{2}[d_{m'n+1}^{j}\left(\beta\right)d_{mn}^{j}\left(\beta\right)\sqrt{\left(j-n\right)\left(j+n+1\right)}\nonumber \\
+d_{m'n-1}^{j}\left(\beta\right)d_{mn}^{j}\left(\beta\right)\sqrt{\left(j+n\right)\left(j-n+1\right)}].
\end{align}

Subtracting Eq. (E4) from Eq. (E3) to eliminate the term $\frac{n\cos\beta}{\sin\beta}d_{mn}^{j}\left(\beta\right)d_{m'n}^{j}\left(\beta\right)$
and multiplying both sides of the equation by $n$, we arrive at a
summation formula 

\begin{align}
\sum_{n}\frac{\left(m-m'\right)n}{\sin\beta}d_{mn}^{j}\left(\beta\right)d_{m'n}^{j}\left(\beta\right)\nonumber \\
=\frac{1}{2}\sum_{n}[d_{m'n+1}^{j}\left(\beta\right)d_{mn}^{j}\left(\beta\right)n\sqrt{\left(j-n\right)\left(j+n+1\right)}\nonumber \\
-d_{mn-1}^{j}\left(\beta\right)d_{m'n}^{j}\left(\beta\right)n\sqrt{\left(j+n\right)\left(j-n+1\right)}\nonumber \\
+d_{m'n-1}^{j}\left(\beta\right)d_{mn}^{j}\left(\beta\right)n\sqrt{\left(j+n\right)\left(j-n+1\right)}\nonumber \\
-d_{mn+1}^{j}\left(\beta\right)d_{m'n}^{j}\left(\beta\right)n\sqrt{\left(j-n\right)\left(j+n+1\right)}].
\end{align}

According to the usual rules of arithmetic, 

\begin{align}
\sum_{n}d_{mn-1}^{j}\left(\beta\right)d_{m'n}^{j}\left(\beta\right)n\sqrt{\left(j+n\right)\left(j-n+1\right)}\nonumber \\
=\sum_{n}d_{mn}^{j}\left(\beta\right)d_{m'n+1}^{j}\left(\beta\right)\left(n+1\right)\sqrt{\left(j-n\right)\left(j+n+1\right)} & ,
\end{align}
and

\begin{eqnarray}
\sum_{n}d_{mn+1}^{j}\left(\beta\right)d_{m'n}^{j}\left(\beta\right)n\sqrt{\left(j-n\right)\left(j+n+1\right)}\nonumber \\
=\sum_{n}d_{mn}^{j}\left(\beta\right)d_{m'n-1}^{j}\left(\beta\right)\left(n-1\right)\sqrt{\left(j+n\right)\left(j-n+1\right)}.\nonumber \\
\end{eqnarray}

As a direct consequence of Eqs. (E5)-(E7), the following invariant
sum is obtained 

\begin{align}
\sum_{n}\frac{\left(m-m'\right)n}{\sin\beta}d_{mn}^{j}\left(\beta\right)d_{m'n}^{j}\left(\beta\right)\nonumber \\
=\frac{1}{2}\sum_{n}d_{mn}^{j}\left(\beta\right)[d_{m'n-1}^{j}\left(\beta\right)\sqrt{\left(j+n\right)\left(j-n+1\right)}\nonumber \\
-d_{m'n+1}^{j}\left(\beta\right)\sqrt{\left(j-n\right)\left(j+n+1\right)}].
\end{align}

\end{document}